\documentclass{amsart}

\usepackage{enumerate}

\numberwithin{equation}{section}
\usepackage{amsmath}
\usepackage{amsthm}
\usepackage{amssymb}

\newcommand{\ess}{\mathrm{ess}}
\newcommand{\BBR}{\mathbb{R}}
\newcommand{\BBE}{\mathbb{E}}

\newcommand{\bE}{\mathbb{E}}
\newcommand{\bR}{\mathbb{R}}

\newcommand{\bN}{\mathbb{N}}

\newcommand{\cB}{\mathcal{B}}
\newcommand{\cF}{\mathcal{F}}
\newcommand{\cG}{\mathcal{G}}
\newcommand{\cK}{\mathcal{K}}
\newcommand{\cM}{\mathcal{M}}

\theoremstyle{plain} \newtheorem{Theo}{Theorem}[section]
\theoremstyle{plain} \newtheorem{Lemma}[Theo]{Lemma}
\theoremstyle{plain} \newtheorem{Cor}[Theo]{Corollary}
\theoremstyle{plain} 
\theoremstyle{remark} \newtheorem{Rem}[Theo]{Remark}
\theoremstyle{definition} 
\theoremstyle{definition} 
\theoremstyle{remark}

\begin{document}

\title[Mutation-selection balance]{Mutation-selection balance with recombination: convergence to equilibrium for polynomial selection costs}

\author{Aubrey Clayton}
\email{aclayton@math.berkeley.edu}
\address{Department of Mathematics \#3840 \\ 
University of California, Berkeley \\
970 Evans Hall \\ 
Berkeley, CA 94720-3840 \\
U.S.A.}

\author{Steven N. Evans}
\email{evans@stat.Berkeley.EDU}
\address{Department of Statistics \#3860 \\
 University of California at Berkeley \\
367 Evans Hall \\
Berkeley, CA 94720-3860 \\
U.S.A}

\thanks{Research supported in part by NSF grant DMS-0405778}

\date{\today}



\begin{abstract}
We study a  continuous-time
dynamical system that models the evolving distribution
of genotypes in an infinite population where
genomes may have infinitely many or even
a continuum of loci, mutations accumulate
along lineages without back-mutation, added
mutations reduce fitness, and recombination occurs
on a faster time scale than mutation and selection.  
Some features of the model, 
such as existence and uniqueness of solutions and convergence to the dynamical
system of an approximating sequence of discrete time models, were presented in 
earlier work by Evans, Steinsaltz, and Wachter for quite general selective costs.  Here we study a special case
where the selective
cost of a genotype with a given accumulation of ancestral
mutations from a wild type ancestor is a sum of costs
attributable to each
individual mutation plus successive interaction
contributions from each $k$-tuple of mutations for $k$
up to some finite ``degree''.
Using ideas from complex chemical reaction networks
and a novel Lyapunov function, 
we establish that the phenomenon of mutation-selection balance
occurs for such selection costs 
under mild conditions.  That is, we show that the dynamical system
has a unique equilibrium and that it converges to this
equilibrium from all initial conditions.
\end{abstract}

\maketitle

\section{Introduction}

Many phenotypic traits, including some
genetic disorders, are thought to be {\em polygenic}
and result from the possibly
complex {\em epistatic} (that is, non-additive)
interactions between large numbers of
mildly deleterious alleles that 
are slowly weeded out of the
population by natural selection
but are constantly reintroduced by recurrent mutation.
In particular, the  Medawar--Williams--Hamilton 
\cite{pM52, gW57, wH66} 
explanation of the evolution of aging
(see, also, \cite{bC94, bC01} and the introductory
discussions in 
\cite{StEvWa05, EvStWa06, 0807.0483, 0808.3622}) invokes
this mechanism of {\em mutation-selection balance}.

It has been a challenge to provide a sufficiently
usable quantitative 
description of the mutation-selection phenomenon 
when many genes are involved (see, for example,
\cite{Hol95} or \cite{FiKi00}: in particular, 
Section 5.3.2 of the latter is entitled ``Need for a mathematical framework''
and specifically lays out the necessity for mathematically
tractable approaches to outcomes that are
modulated by the interaction of large numbers of genes).

%Two opposing processes in population dynamics are natural selection and recurring deleterious mutation.  Specifically, the genetic makeup of a population is the result of a balance between the continual introduction of genetic mutations with an adverse effect on fitness and the selective process of weeding these mutations out.  
Evans, Steinsaltz, and Wachter \cite{StEvWa05, EvStWa06}
describe two related mathematical settings
for the mutation-selection process.
Here we consider the model of the latter paper.  
The key assumptions in \cite{EvStWa06} are:
\begin{itemize}
\item
the population is infinite,
\item
the genome may consist of infinitely many 
or even a continuum of loci, 
\item
reproduction is sexual, in that each individual has
two parents and the mechanism of {\em genetic recombination}
randomly shuffles together the genomes of the parents
in order to obtain the genome of the offspring,
\item
mating is random,
\item
individuals are {\em haploid} (that is,
an individual has only one copy of each gene
rather than copies from each of
two parents, as is the case for {\em diploid}
organisms),
\item
mutations accumulate
along a lineage from an ancestral {\em wild type}
genotype
(that is, there is no back-mutation canceling out
the effect of earlier mutations),
\item 
fitness is calculated for
individuals rather than for mating pairs,
\item
genotypes that
have more accumulated mutations are less fit,
but otherwise selection costs are arbitrary,
\item
the effects of recombination are felt on a faster
time scale than those of mutation or selection.
\end{itemize}

For the purpose of explaining our results, we now describe briefly the general model of \cite{EvStWa06}.  Our
results concern a particular family of
selection costs that we introduce later.

We write $\cM$ for the collection of {\em loci}
in the portion of the genome that is of interest to us.
We assume that there is a distinguished reference
wild type genotype, and each locus represents
a ``position'' at which the genotype of an individual
may differ from that of the wild genotype.  
We have placed the word ``position'' in quotes, because we 
do not necessarily take $\cM$ to be something like
a finite collection of physical DNA base positions
or a finite collection of genes.
From a modeling point
of view, it is convenient to allow $\cM$ to be more
general.  For example, the
proposed explanations in Charlesworth \cite{bC01} 
for Gompertz mortality curves and mortality plateaus
at extreme ages suggest that one relevant
choice of $\cM$ may be a class of functions from
$\bR_+$ to $\bR_+$: the value of such a function
at time $t \ge 0$ represents an additional increment
to the mortality hazard rate at age $t$ conferred
by a mutation away from the wild type at this locus.
Some minimal amount of structure on $\cM$ is necessary in order
to accommodate probability theory, and so we
take $\cM$ to be a complete,
separable metric space.

We suppose that the genotype of an individual
is described completely by the number of times a mutation
event has occurred at each locus along the lineage
connecting the individual to a wild type ancestor.
Similar assumptions are standard in models
of the {\em infinitely many alleles}, {\em
infinitely many sites}, and {\em continuum-of-alleles} type.
In particular, all potential
mutations at a locus
that is currently wild type are indistinguishable, 
and further
mutation events cannot undo the results of earlier ones
(that is, there is no {\em back mutation}). 
When the space $\cM$ is continuous, it will
typically be the case that multiple mutations
will not occur along a lineage at any locus. For 
discrete spaces $\mathcal{M}$ 
that are large, the assumption of mutation
accumulation without back-mutation
is reasonable when mutation rates are 
sufficiently low that it is unlikely that any locus
will be affected by a mutation more than once along
a lineage.

Consequently, we may think of 
an individual's genotype as an element of the
space $\cG$ of
integer--valued finite Borel measures on $\cM$,
where the non-negative integer $g(A)$ for
$g \in \cG$ and $A \subseteq \cM$ represents
the total number of mutation events that
have occurred at loci belonging to the subset
$A$ of the genome along the lineage connecting
the individual to a wild type ancestor.
An element of $\cG$ is a finite sum $\sum_i \delta_{m_{i}}$,
where $\delta_m$ is the unit point mass at
the locus $m \in \cM$, corresponding to a genotype
where there have been ancestral
mutations at loci $m_1, m_2, \ldots$.  In
particular, the wild type
genotype is the null measure.  As we remarked above, we do not
require that the loci
$m_{i} \in \cM$  are distinct. For example,
if $\mathcal{M}$ is finite, so we might as well
take $\mathcal{M} = \{1,2,\ldots,N\}$ for some positive
integer $N$, then a
genotype is of the form $\sum_{j=1}^N n_j \delta_j$,
indicating that ancestral mutations have
occurred $n_j$ times at locus $j$.  We may identify
such a genotype with the
non-negative integer vector $\mathbf{n} := (n_1, n_2, \ldots, n_N)$,
so that $\mathcal{G}$ is essentially
the Cartesian product $\mathbb{N}^N$ of $N$ copies
the non-negative integers.

We imagine that the population we are interested in
is infinite and that all that matters about
an individual is the individual's genotype, so that
the dynamics of the population can be described in
terms of the proportions of individuals with
genotypes belonging to the various subsets of $\cG$.
We therefore seek to model the evolution
of probability measures $P_t$, $t \ge 0$, on $\cG$, 
where $P_t(G)$ represents the proportion of individuals
in the population at time $t$ that have genotypes
belonging to the set $G \subseteq \cG$.
Equivalently, $P_t(G)$ can be interpreted as the
probability that an individual chosen uniformly at random from
the population will have genotype belonging to $G$,
so that $P_t$ is the distribution of
a random finite integer-valued measure on $\cM$.
For example, if $\cM = \{1,2,\ldots,N\}$
and we identify $\cG$ with $\bN^N$ as above, then
$P_t(\{\mathbf{n}\})$ represents the
probability that an individual chosen uniformly at random
from the population will have $n_j$ ancestral mutations
at locus $j$ for $j=1, 2,\ldots,N$.

We remark that although the introduction of
this measure-theoretic formalism might
seem somewhat heavy-handed, the use of measures
to represent genotypes and probability measures on
spaces of measures to represent populations
 has become quite standard in mathematical
population genetics.  It is apparent
from \cite{MR838085} how useful this perspective 
has been for understanding stochastic models of the 
infinitely-many-alleles and infinitely-many-sites type.
Similarly, the stochastic model of \cite{SaHa92} is
based on a Poisson random measure on a Cartesian product of
the space of sites and the unit interval (with the latter coordinate
representing frequency of mutant alleles at the corresponding site).
Also, the recent papers
\cite{MR1952324,MR2191729, MR2297255}  adopt a measure-theoretic framework
to present a deterministic model of recombination based on \cite{MR1842838}.  The discrete-time
mutation-selection models of \cite{MR0356905} and
\cite{MR0465272} describe
the population in terms of a probability measure over
the real numbers that records the proportions of
individuals with various fitnesses.  A far-reaching
continuous-time generalization of these models is presented in \cite{MR1372337} (see also \cite{MR1885085})
where individuals have an abstract {\em type}
belonging to a locally compact space and the population
is described by a probability measure on this space.

We next indicate how mutation, selection
and recombination are modeled in our setting
to obtain the evolution dynamics for $P_t$.

\bigskip\noindent
{\bf Mutation alone.}  Suppose that there is only mutation
and no selection or recombination.  Because no
genome has a selective advantage, all
individuals are dying and reproducing at the same
rate, and since there is no recombination the
genotype of an offspring is simply that of the
parent with possible alterations due to
mutation.  Somewhat informally,
we imagine that there is finite measure
$\nu$ on the space of loci $\cM$ such that
for any individual
alive at time $t \ge 0$ the probability
a mutation occurs in the infinitesimal region
$dm \subset \cM$ during the time interval $dt$ is
$dt \times \nu(dm)$, and we also assume that
the occurrences of mutations are independent
between different individuals, different times,
and different regions of $\cM$.
If we write
$P_t \Phi = \int_{\cG} \Phi(g) P_t(dg)$ for some
test function $\Phi: \cG \rightarrow \bR$ (that is,
$P_t \Phi$ is the expected value of the real-valued
random variable
obtained by applying the function $\Phi$ to the genotype
of an individual chosen uniformly at random from the population),
then the content of
these assumptions is contained formally in the equation
\begin{equation*} 
   \frac{d}{dt}P_{t}\Phi
      = P_{t} \left( \int_\cM \left[ \Phi(\cdot+\delta_{m}) -
      \Phi(\cdot) \right] \, \nu(dm) \right).
\end{equation*}

For example, when $\cM = \{1,2,\ldots,N\}$ we have
\begin{equation*}
\frac{d}{dt} P_t(\{\mathbf{n}\}) \\
=
\sum_{j=1}^N
\nu(\{j\}) 
\left[
P_t(\{\mathbf{n} - \mathbf{e_j}\}) - P_t(\{\mathbf{n}\})
\right],
\end{equation*}
where $\mathbf{e_j}$ 
is the $j^{\mathrm{th}}$ coordinate vector. 
%\begin{equation*}
%\begin{split}
%& \frac{d}{dt} P_t(\{(n_1, n_2, \ldots, n_N)\}) \\
%& \quad =
%\sum_{j=1}^N
%\nu(\{j\}) 
%\left[
%P_t(\{(n_1, n_2, \ldots, n_j - 1, \ldots, n_N)\}) - %P_t(\{(n_1, n_2, \ldots, n_N)\})
%\right]. \\
%\end{split}
%\end{equation*}
We stress that this equation is classical: it is a special
case of the usual equation describing evolution due to mutation
of type frequencies in a population where the set
of types is $\bN^N$ and mutation from
type $\mathbf{n}$ to type 
$\mathbf{n} + \mathbf{e_j}$ occurs at
rate $\nu(\{j\})$ -- see, for example, Section III.1.2
of \cite{MR1885085}.

This evolution equation has a simple explicit solution.
Let $\Pi$ denote a Poisson random 
measure on $\cM \times \bR_+$
with intensity measure
$\nu \otimes \lambda$, where $\lambda$
is Lebesgue measure; that is,
$\Pi$ is a random integer-valued Borel measure
such that:
\begin{itemize}
\item[(1)] 
The random variable $\Pi(A)$ is Poisson with expectation
$\nu \otimes \lambda(A)$ for any Borel subset $A$ of $\cM \times \bR_+$.
\item[(2)] 
If $A_1, A_2, \ldots, A_n$ are disjoint Borel subsets of 
$\cM \times \bR_+$, then the random variables $\Pi(A_k)$ are independent.
\end{itemize}
Define a $\cG$-valued random variable
$Z_t$ (that is, $Z_t$ is a random finite
integer-valued measure on $\cM$) by 
$
Z_t := \int_{ \cM \times [0,t]} \delta_m \, \Pi(d(m,u)).
$
Then
$
P_t \Phi = {\bE} \left[\Phi(W+Z_t)\right]
$,
where $W$ is a random measure on $\cM$
that is independent of $\Pi$
and $W$ has distribution $P_{0}$.
In particular, if $P_0$ is the distribution
of a Poisson random measure, then $P_t$ will
also be the distribution of Poisson random measure.
If we write
$\rho_t$ for the intensity measure associated with $P_t$
(that is, $\rho_t$ is the measure on
$\cM$ defined by $\rho_t(M) := \int_\cG g(M) \, P_t(dg)$
for $M \subseteq \cM$), then
$\rho_t$ evolves according to
the simple dynamics
$\rho_t(M) = \rho_0(M) + t \, \nu(M)
$.

\bigskip\noindent
{\bf Selection alone.} Now consider what happens if there
is only selection and no mutation or recombination.
We assume that there is a
{\em selection cost function} $S : \cG \to \BBR_+$ with
the interpretation that $S(g') - S(g'')$ for
$g', g'' \in \cG$ represents
the difference in the rate of population size increase
between individuals with genotype $g''$
and individuals with genotype $g'$.
We make the normalizing assumption
$S(0) = 0$ and suppose that
\begin{equation}
\label{monotone_cost}
S(g+h)\geq S(h), \quad g,h \in \cG,
\end{equation} 
so that if the ancestral mutations from wild type
in one genotype are a subset (taking into account multiplicity)
of those in a second genotype, then the individuals
with the second genotype have a lesser rate of increase
than those with the first genotype.  
Heuristically, we have
\begin{equation*} 
   \frac{d}{dt}P_{t}(dg')
   =
   \frac{d}{du} 
   \frac{\exp(-S(g') u) P_t(dg') }
   {\int_{\cG} \exp(-S(g'') u) \, P_t(dg'')}
   \Big |_{u = 0}
   = (P_t S - S(g')) \, P_t(dg');
\end{equation*}
That is, at time $t \ge 0$ the per individual rate 
of increase of the proportion of the population
of individuals with genotype $g'$
is $P_t S - S(g')$.  More formally, we have
\begin{equation*} 
   \frac{d}{dt}P_{t}\Phi
      = - \, P_{t}(\Phi \cdot [S -P_{t}S])
      = - \int_\cG \Phi(g')
      \left[S(g') - \int_\cG S(g'') \, P_t(dg'')\right] \, P_t(dg').
\end{equation*}
For example, when $\cM = \{1,2,\ldots,N\}$ we have
\begin{equation*}
\frac{d}{dt} P_t(\{\mathbf{n'}\})
=
-\left[ 
S(\mathbf{n'})
-
\sum_{\mathbf{n''}} P_t(\{\mathbf{n''}\}) S(\mathbf{n''})
\right]
P_t(\{\mathbf{n'}\}).
\end{equation*}
%\begin{equation*}
%\begin{split}
%& \frac{d}{dt} P_t(\{(n_1', n_2', \ldots, n_N')\}) \\
%& \quad =
%-\left[ S(n_1', n_2', \ldots, n_N')
%-
%\sum_{(n_1'', n_2'', \ldots, n_N'')} P_t(\{(n_1'', n_2'', \ldots, n_N'')\}) S(n_1'', n_2'', \ldots, n_N'') \right] \\
%& \qquad \times P_t(\{(n_1', n_2', \ldots, n_N')\}). \\
%\end{split}
%\end{equation*}
We again stress that this equation is classical: it is a special
case of the usual 
equation describing evolution due to selection
of type frequencies in a population where the set
of types is $\bN^N$ -- see, for example, Section III.1.2
of \cite{MR1885085}.

If $S$ is {\em non-epistatic} (that is,
$S$ has the additive property 
$S(\sum_i \delta_{m_{i}}) = \sum_i S(\delta_{m_{i}})$),
then there is no
interaction between the selective effects of ancestral mutations
at different loci or multiple ancestral mutations
at a single locus:  in particular, if $P_0$ is the
distribution of a Poisson random measure then
$P_t$ will also be the distribution of a Poisson
random measure and, writing $\rho_t$ for the
intensity measure associated with $P_t$ as before, we have
\begin{equation*}
\rho_t(dm') = \rho_0(dm')
- \int_0^t
\left[S(\delta_{m'}) - \int_{\cM} S(\delta_{m''}) \rho_s(dm'')\right] \,
\rho_s(dm') \, ds. 
\end{equation*}
However, if $S$ is epistatic (that is, non-additive),
then $P_t$ will, in general, not be the distribution
of a Poisson random measure -- even when $P_0$ is.

\bigskip\noindent
{\bf Recombination alone.} Lastly, we discuss the incorporation of recombination.  Our
description will be brief, as the details will not be important 
to us, because we are interested in an asymptotic regime
where recombination occurs on a faster time-scale
than mutation or selection, and in the limit
the detailed features of the recombination mechanism
disappear.  
The effect of recombination is to choose an individual
uniformly at random from the population at some rate and replace the
individual's genotype $g'$ with a genotype of the form
$g'(\cdot \cap M) + g''(\cdot \cap M^c)$, where $g''$ is the
genotype of another randomly chosen individual, $M$ is a subset
of $\cM$ chosen according to a suitable random mechanism,
and $M^c$ is the complement of $M$.  Thus, recombination
randomly shuffles together two different genotypes drawn
from the population. If 
recombination alone is acting, $P_0$ has the property that
there does not exist an $m \in \cM$ with $P_0(\{g \in \cG : g(\{m\}) > 0\})$,
and the mechanism for choosing the so-called
{\em segregating} set $M$
is such that, loosely speaking, if $m'$ and $m''$ are
two loci then there is positive probability that
$m' \in M$ and $m'' \in M^c$, so that no
region of $\cM$ is immune from shuffling,
then $P_t$ will converge to a Poisson random measure 
with the same intensity measure as $P_0$ as
$t$ increases -- see \cite{EvStWa06} for details.

\bigskip\noindent
{\bf Combining mutation, selection and recombination.}
We have seen that if $P_0$ is the distribution of a Poisson
random measure, then mutation preserves this property, while
(epistatic = non-additive) selection and
recombination respectively drive the system away from and toward
Poisson.  If all three processes are operating and we consider
a limiting regime where recombination acts on a faster time scale
than mutation and selection, then we expect that
the resulting system will preserve the Poisson property.  Results
to this effect are established in \cite{EvStWa06}: more
precisely, it is shown there that if one considers
a discrete generation evolution in which mutation
and selection are modeled by discrete-time
analogues of the differential equations described
above, recombination is modeled as above, and time
is scaled so that the time between generations
converges to zero, then the  discrete-time evolution
converges to a continuous-time evolution
that preserves Poisson initial conditions, provided
that asymptotically recombination acts on
a faster time scale than mutation and selection.
Moreover, the explicit details of the recombination
mechanism in the discrete generation
evolutions have no influence on the limit:  all that
matters is that the relative effect of recombination becomes
stronger and stronger in the limit, and,
as described above,  that 
no region of $\cM$ is immune from the shuffling
effect of recombination.
We refer the reader
to \cite{EvStWa06} for rigorous statements and proofs;
however, we do stress that the analysis of
\cite{EvStWa06} does {\bf not} start from an
{\em a priori} assumption that loci are
unlinked, rather it delineates how strong
the Poissonizing effect of recombination has
to be in order to overcome the tendency of
non-epistatic selection to induce linkage
between loci.

The limiting evolution is thus
a family of probability measures $P_t$ on $\cG$
that are Poisson with intensity measure $\rho_t$.  If we write
$X^\pi$ for a Poisson random measure on $\cM$ with intensity measure
$\pi$, then, as we expect from combining the observations
above, $\rho_t$ satisfies the evolution equation
\begin{equation}
\label{model_informal}
\rho_t(dm) = \rho_0(dm) + t\, \nu(dm) - 
\int_0^t  
\bE\left[S(X^{\rho_s} + \delta_m) - S(X^{\rho_s})\right] 
\, \rho_s(dm) \, ds.
\end{equation} 
For example, when $\cM = \{1,2,\ldots,N\}$ we have
\begin{equation*}
\frac{d}{dt} \rho_t(\{j\})
=
\nu(\{j\})
-
\rho_t(\{j\}) \sum_{\mathbf{n}} 
\left[
S(\mathbf{n}+\mathbf{e_j}) - S(\mathbf{n})
\right]
\prod_{k=1}^N \exp(-\rho_t(\{k\})) \frac{\rho_t(\{k\})^{n_k}}{n_k!}.
\end{equation*}
%\begin{equation*}
%\begin{split}
%& \frac{d}{dt} \rho_t(\{j\}) \\
%& \quad =
%\nu(\{j\}) \\
%& \qquad -
%\rho_t(\{j\}) \sum_{(n_1, n_2, \ldots, n_N)} 
%\left[
%S((n_1, n_2, \ldots, n_j + 1, \ldots, n_N)) - S((n_1, n_2, \ldots, n_N))
%\right] \\
%& \quad \qquad \times
%\prod_{k=1}^N \exp(-\rho_t(\{k\})) \frac{\rho_t(\{k\})^{n_k}}{n_k!}. \\
%\end{split}
%\end{equation*}
Note that although there are infinitely many ``types''
in this latter finitely-many-loci special case, 
the evolution of the population is described 
by a system of $N$ differential equations 
for the $N$ real quantities $\rho_t(\{j\})$,
$1 \le j \le N$.

In this paper we consider selection costs $S$ that are, in
a suitable sense, {\em polynomial}.  Roughly
speaking, this means that, 
for some finite positive integer $N$,
the selection cost
of a given genotype (that is, a given collection
of accumulated mutations from the ancestral wild type) is
a sum of non-negative contributions from
each individual mutation, plus a sum of non-negative
contributions due to interactions between the mutations
in every pair
of mutations, and so on -- all they way up to
a sum of non-negative contributions due to
interactions between the mutations in every
$N$-tuple of mutations.

We show for such selection costs that, under very
mild conditions, the system (\ref{model_informal}) has
a unique equilibrium and that the system converges to
this equilibrium from any initial condition $\rho_0$
that is absolutely continuous with respect to $\nu$.

Because defining precisely
what we mean by a polynomial selection
cost and describing our results
for a general locus space $\cM$ requires a certain amount
of extra notation, for the sake of this introduction we just
describe our results in the special case when
$\cM$ is the finite set $\{1,...,N\}$, and leave the
general case to the body of the paper.

In the finitely-many-loci case, the mutation rate
measure $\nu$ is defined by its value on singleton sets
and we write $\nu_i := \nu (\{i\})$.  
We assume that  $\nu_i > 0$ for all $i$, and let 
$\nu$ also denote the vector $(\nu_1,\ldots ,\nu_N)$.  As we remarked above,
a genotype $g \in \cG$ can  be encoded by an ordered $N-$tuple of non-negative integers 
$g=(g_1,\ldots,g_N)$, where $g_k$ represents the number of ancestral mutations that are present at locus $k$.  
A polynomial selection cost is one of the form 
\begin{equation}
\label{finite_poly_selection}
S(g)=\sum_{I} \alpha_I g^I, 
\end{equation}
where the sum is taken over all nonempty subsets $I \subseteq \{1,\ldots,N\}$
and we adopt the convention that for a vector 
$v$ the notation $v^I$ denotes the product $\prod_{i \in I} v_i$.
The constants $\alpha_{\{i\}}$ for $1 \le i \le N$ measure the selective
cost of a mutation at locus $i$ alone, whereas the constants $\alpha_I$ for subsets $I \subseteq \{1,\ldots,N\}$ 
of cardinality greater than one measure the 
selective cost attributable to interactions 
between the mutations at loci
in $I$ over and above that attributable to interactions between
mutations in sets of loci that are proper subsets of $I$.  
We assume that $\alpha_I \ge 0$ for all $I$. 
This assumption ensures that the monotonicity condition (\ref{monotone_cost})
holds. It is, of course, not a necessary condition for (\ref{monotone_cost}) to hold,
but it will be crucial in our analysis.  Indeed, it
will be shown in a forthcoming paper by Evans, Steinsaltz and Wachter
that there are some natural infinite dimensional
systems that arise in applications to aging where the analogue of 
(\ref{finite_poly_selection}) fails to hold and the resulting system does not converge
to an equilibrium.

Write $(\rho_1(t),...,\rho_N(t))$ for the vector corresponding
to the intensity measure of $P_t$.  As we show in Section~\ref{S:model},
the evolution equation (\ref{model_informal}) is equivalent to
the system of ordinary differential equations
\begin{equation}
\label{finitesysdiff}
\dot{\rho_k} = \nu_k - \sum_{I \in \mathcal{I}_k}  \, \alpha_I \rho^I, \quad 1 \le k \le N,
\end{equation}
where $\mathcal{I}_k$ denotes the collection of 
subsets of $\{1,\ldots,N\}$ that contain $k$.  
The following is an immediate corollary
of our main result for general locus spaces $\cM$, Theorem~\ref{stability}.

\begin{Cor}
\label{exmprop}
Suppose that $\alpha_{\{i\}}>0$ for $1 \le i \le N$.
The system (\ref{finitesysdiff}) has a unique equilibrium point in the positive orthant
$\bR_+^N$, and this equilibrium is globally stable (that is, the system converges
to the equilibrium from any initial conditions in $\bR_+^N$).
\end{Cor}

Note that, in this finitely-many-loci case, 
the problem of determining the equilibrium point
reduces to finding the unique zero in the positive
orthant of a set of $N$ equations in $N$ variables,
and there is an extensive body of theory in numerical
analysis and computational commutative algebra devoted
to this problem.  

We finish this introduction with a very 
brief indication of the substantial literature on multilocus deterministic models in population genetics 
and the particular import of such models for the
phenomenon of mutation-selection balance.

The standard mathematical reference on population genetics models in general
(both deterministic and stochastic) is 
\cite{MR2026891} by Warren Ewens. This is a revision of a book
from 1979 and the treatment of deterministic models is
not expanded significantly  from the original.  
In \cite{MR0682090}, Ewens and Glenys Thomson also
made one of the earliest
contributions to understanding equilibria of deterministic multilocus genetic systems in which no
special assumptions are made about fitness structure.
While mutation only occurs from the wild type
to the derived type in our model, the
models in \cite{MR0682090} (in common with many other models in the literature)
allow several alleles at a locus and mutation from
any allelic type to any other.  

A very comprehensive recent reference that
concentrates on the infinite-population,
deterministic aspects of population genetics
is Reinhard B\"urger's book \cite{MR1885085} (see also the B\"urger's review paper \cite{Bur98}).  
In particular, these works consider at 
length deterministic haploid {\em continuum-of-alleles}
models in which individuals have a type which is
thought of as the contribution of a gene to a given
quantitative trait.  The type belongs to a general state space that represents something like
the trait value (in which case the state space is a subset of $\bR$) and types
have an associated fitness, which is some fairly arbitrary function from the type space to
$(0,1]$.  The type  may be regarded as the combined effect of a multilocus genotype,
but the models do not explicitly incorporate a family of loci, the configuration of alleles
present at those loci, or a function describing the fitness of a configuration: everything
is cast in terms of how fit each type is and how likely one type is to mutate into another.

The development in \cite{EvStWa06} leading to the model we consider here
was very much influenced by the discrete time multilocus models of \cite{TuBa90, BaTa91, KiJoBa02}
that incorporate arbitrary forms of selection,
general modes of inheritance, and other evolutionary forces such as migration and mutation.
Also, the model studied here and in \cite{EvStWa06} is essentially the result of adding strong recombination to
the model of
\cite{StEvWa05}, which may itself be thought of as a
generalization of the {\em infinitesimally rare alleles} model
of \cite{KiMa66} (see also, \cite{Kon82})
as it is cast in \cite{Daw99}.

Certain classes of mutation-selection models without recombination 
are solved explicitly
in \cite{MR1657856, BaWa01} using ideas from statistical mechanics.  
Such models may be thought of as either
multilocus systems with complete linkage or structured single locus systems.
Finally, we have already mentioned the constellation of papers 
\cite{MR1952324,MR2191729, MR2297255, MR1842838} presenting a deterministic
model of population change due to recombination alone.

\section{The Model}
\label{S:model}

In this section, we review the details of the model in (\cite{EvStWa06}) and derive the relevant equations in the special case of polynomial selection costs, to be defined below.

Let $\cM$, $\cG$, $\nu$, and $S$ be as above.  That is,
$\cM$ is the complete, separable metric space of loci, 
$\cG$ is the space of integer-valued finite measures on $\cM$
representing possible genotypes,
$\nu$ is the finite measure on $\cM$ describing the rates at which mutation events occur in different parts of the genome, and
$S: \cG \rightarrow \bR_+$ is the selection cost function.  As above, we assume that
$S(0) = 0$ and $S(g+h) \ge S(g)$ for $g,h \in \cG$.

For any finite measure $\pi$ on $\cM$, let $X^\pi$ denote a Poisson random measure on 
$\cM$ with intensity $\pi$ (so that $X^\pi$ is a random variable
with values in $\cG$).  Define a function $F_\pi : \cM \to \BBR_+$ by 
\begin{equation}
F_\pi(m) = \BBE [S(X^\pi + \delta_m) - S(X^\pi)]
\end{equation}
That is, $F_\pi(m)$ measures the average change in selection cost when mutation at locus 
$m \in \cM$ is added to the random genotype $X^\pi$.

Let $D \pi $ be the measure on $\cM$, absolutely continuous with respect to $\pi$, 
with density (that is, Radon-Nikodym derivative) $F_\pi$: 
\begin{equation}
\frac{d (D \pi) }{d \pi} (m) = F_\pi (m).
\end{equation}

In this notation, equation (\ref{model_informal}) becomes
\begin{equation}
\label{model}
\rho_t = \rho_0 + t \nu - \int_0^t D \rho_s \, ds,
\end{equation}
where, as above, $\rho_t$ is the finite measure on $\cM$ giving
the intensity measure of the Poisson random measure with
distribution $P_t$.  We may write (\ref{model}) somewhat informally as
\[
\rho_t(dm) = \rho_0(dm) + t \nu(dm) - \int_0^t F_{\rho_s}(m) \rho_s(dm) \, ds.
\]

It was shown in \cite{EvStWa06} that (\ref{model}) has a unique solution
for any initial condition $\rho_0$ provided that the selection cost 
$S$ satisfies a Lipschitz condition with respect to the Wasserstein metric
on $\cG$.  We take a slightly different perspective here.
Let $L^\infty(\cM, \nu)$ denote the usual
Banach space  of (equivalence classes of)
$\nu$-essentially bounded 
functions on $\cM$, and write
$L_+^\infty(\cM, \nu)$ for the subset of $L^\infty(\cM, \nu)$
 consisting of non-negative functions.  There is, of course,
a bijection between  $L_+^\infty(\cM, \nu)$ and the space $\cK$ of
finite measures on $\cM$ that are absolutely continuous with
respect to $\nu$ with $\nu$-essentially bounded Radon-Nikodym derivative,
and so we can metrize $\cK$ using the $L^\infty(\cM, \nu)$ metric
on $L_+^\infty(\cM, \nu)$.  An argument essentially the same as that
in \cite{EvStWa06} establishes the following.

\begin{Lemma}
\label{L:exist_unique}
Suppose that $S$ is such that 
$F_\pi \in L_+^\infty(M,\nu)$ for $\pi \in \cK$ and
there exists
a function $H:\bR_+ \rightarrow \bR_+$ for which
\[
\left\| 
\frac{d (D \pi')}{d \nu} - \frac{d (D \pi'')}{d \nu}
\right\|_\infty
=
\|F_{\pi'} \phi' - F_{\pi''} \phi''\|_\infty 
\le H(\|\phi'\|_\infty \vee \|\phi''\|_\infty) \|\phi' - \phi''\|_\infty
\]
when $\pi', \pi'' \in \cK$ with
$\pi'(dm) = \phi'(m) \, \nu(dm)$ and $\pi''(dm) = \phi''(m) \, \nu(dm)$.
Then (\ref{model}) has a unique 
$\cK$-valued solution for any $\rho_0 \in \cK$.
\end{Lemma}

From now on we will assume that the selection cost $S$
is {\em polynomial}, by which we mean that
\begin{equation}
\label{general_poly_selection}
S(g)=\sum_{n=1}^N \int_{\cM^n} a_n (\mathbf{m}) \, g^{\otimes n} (d\mathbf{m})
\end{equation}
for some positive integer $N$,
where for each $n$ the Borel function $a_n : \cM^n \to \mathbb{R}_+$ is permutation-invariant
 (that is, $a_n(\pi \mathbf{m})=a_n(\mathbf{m})$ for all permutations $\pi$) 
and  has the property that $a_n(\mathbf{m})=0$ if there exist $i \neq j$ with $m_i = m_j$.  
Furthermore, we assume that each function
$a_n$ is bounded.  The number $n! \, a_n(m_1,\ldots,m_n)$ 
represents the interactive effect of possessing mutations
at the $n$ loci $m_1,\ldots,m_n$ 
over and above that of possessing any subset of them, and
this additional effect is independent of the order in which the mutations are written.

\begin{Rem}
When $\cM$ is the finite set $\{1,...,N\}$ and, as above, we encode
a genotype $g \in \cG$ as an ordered $N-$tuple of non-negative integers 
$g=(g_1,\ldots,g_N)$, where $g_k$ represents the number of times
an ancestral  mutation has occurred at locus $k$, then
the expression for $S(g)$ in (\ref{general_poly_selection}) coincides with that in 
(\ref{finite_poly_selection}) if we set $\alpha_I = n! \, a_n(\mathbf{m})$ when 
$I$ is the subset $\{m_1, \ldots, m_n\} \subseteq \{1,\ldots,N\} = \cM$ and 
$\mathbf{m}$ is the vector $(m_1, \ldots, m_n)$. 
\end{Rem}

%\begin{Rem}
%As we remarked in the Introduction for the special case when 
%$\cM$ is finite, the 
%assumption that each $a_n$ is non-negative is sufficient but certainly not
%necessary for (\ref{monotone_cost}) to hold.  However, it is central
%to our analysis.
%\end{Rem}

%The connection with ordinary polynomials will be made clear in Example (\ref{finiteexm}).

We will now derive the particular form that the equation (\ref{model}) takes
for polynomial selections costs.
Fix $\pi \in \cK$ and, as above, let $X^\pi$ be a Poisson random measure with intensity $\pi$.  
Then, for each $m \in \cM$, we have 
\begin{equation*}
\begin{split}
S(X^\pi + \delta_m) 
& = 
\sum_{n=1}^N \int_{\cM^n} a_n (\mathbf{m}) (X^\pi + \delta_m)^{\otimes n} (d\mathbf{m}) \\
& =
\sum_{n=1}^N \int_{\cM^n} a_n(\mathbf{m}) 
\left ( \sum_{k=0}^n 
\binom{n}{k} (X^\pi)^{\otimes k} \otimes \delta_m ^{\otimes (n-k)} \right) (d\mathbf{m}). \\
\end{split}
\end{equation*}
Note that the symmetry of $a_n$ allows us to rearrange the order of variables in the integral.
Thus,
\begin{equation*}
\begin{split}
& S(X^\pi + \delta_m) 
 = 
\sum_{n=1}^N \sum_{k=0}^n \binom{n}{k} 
\int_{\cM^n} a_n(\mathbf{m}) \, (X^\pi)^{\otimes k} \otimes \delta_m ^{\otimes (n-k)}(d\mathbf{m}) \\ 
& \quad = 
\sum_{n=1}^N \left [ \int_{\cM^n} a_n(\mathbf{m}) (X^\pi)^{\otimes n}(d\mathbf{m}) + n \int_{\cM^n} a_n(\mathbf{m}) \, (X^\pi)^{\otimes (n-1)} \otimes \delta_m (d\mathbf{m}) \right], \\
\end{split}
\end{equation*}
because all other integrals are 0 by the assumptions on $a_n$.
Hence, 
\begin{equation}
S(X^\pi + \delta_m) 
=S(X^\pi) + \sum_{n=1}^N n \int_{\cM^{(n-1)}} a_n(\mathbf{m},m) \, (X^\pi)^{\otimes(n-1)} (d \mathbf{m}),
\end{equation}
where we write $a_n(\mathbf{m},m) := a_n(m_1, \ldots, m_{n-1},m)$ for 
$\mathbf{m} = (m_1, \ldots, m_{n-1}) \in \cM^{(n-1)}$
and $m \in \cM$, and 
by convention, $\int_{\cM^0} a_1(\mathbf{m},m) \, (X^\pi)^{\otimes(0)} (d \mathbf{m})=a_1(m)$.
Therefore, 
\begin{equation}
\label{frho}
F_\pi (m) = \mathbb{E} \left [S(X^\pi + \delta_m) - S(X^\pi) \right] = \sum_{n=1}^N n \int_{\cM^{(n-1)}} a_n(\mathbf{m},m) \, \pi^{\otimes(n-1)} (d \mathbf{m}),
\end{equation}
by the independent increments property of Poisson random measures.

\begin{Rem}
Suppose that $\cM$ is the finite set $\{1,...,N\}$ and, as above, we encode
a genotype $g \in \cG$ as an ordered $N-$tuple of non-negative integers 
$g=(g_1,\ldots,g_N)$ and set $\alpha_I = n! \, a_n(\mathbf{m})$ when 
$I$ is the subset $\{m_1, \ldots, m_n\} \subseteq \{1,\ldots,N\} = \cM$ and 
$\mathbf{m}$ is the vector $(m_1, \ldots, m_n)$.  Then 
\[
F_\pi(k) = \sum_{I \in \mathcal{I}_k} \alpha_I \pi^I, \quad k \in \cM,
\]
where we identify the measure $\pi$ on $\cM$ with the vector $(\pi_1, \ldots, \pi_N)$
and we use the notations of the Introduction that $\pi^I = \prod_{i \in I} \pi_i$
and $\mathcal{I}_k$ is the collection of subsets of $\cM$ that contain $k$.
Consequently, equation (\ref{model}) becomes equation (\ref{finitesysdiff})
in this special case.
\end{Rem}

Returning to the general case,
it is straightforward to check that the condition of Lemma~\ref{L:exist_unique} holds
for our choice of the cost function
$S$, and so (\ref{model}) has a unique $\cK$-valued solution for any 
initial condition $\rho_0 \in \cK$.  We will henceforth
suppose that $\rho_0 \in \cK$.

Write $\phi_t$ for a choice of the Radon-Nikodym derivative
of $\rho_t$ with respect to $\nu$.  It will be convenient
to turn our measure-valued
integral equation (\ref{model}) for $t \mapsto \rho_t$
into a function-valued differential equation for the
corresponding Radon-Nikodym derivatives with respect to
$\nu$ as follows.

For $x \in L_+^\infty(\cM,\nu)$, define 
$G x \in L_+^\infty(\cM,\nu)$ by
\[
G x(m) := 
\left[
\sum_{n=1}^N n \int_{\cM^{(n-1)}} a_n(\mathbf{m},m) x(m_1)\cdots x(m_{n-1}) \, \nu ^{\otimes (n-1)}(d\mathbf{m})
\right]
x(m).
\]

\begin{Lemma}
Suppose that $\rho_t$ is the solution of (\ref{model}).
Then there is a non-negative Borel function 
$(t,m) \mapsto x_t(m)$ 
on $\bR_+ \times \cM$
such that the function
$m \mapsto x_t(m)$ is a Radon-Nikodym
derivative of $\rho_t$ with respect to $\nu$
for all $t \ge 0$, and for $\nu$-a.e. $m \in \cM$ the function 
$t \mapsto x_t(m)$ is differentiable with 
\begin{equation}
\label{sys}
\dot{x}_t(m) = 1 - G x_t(m),
\end{equation}
\end{Lemma}

\begin{proof}
For each $t \ge 0$, let $\phi_t$ 
be a Radon-Nikodym
derivative of $\rho_t$ with respect to $\nu$.
We may write $\cB(\cM)$, the Borel $\sigma$-field
on $\cM$, as $\cB(\cM) = \sigma(\bigcup_k \cF_k)$,
where $\cF_1 \subseteq \cF_2 \subseteq \ldots$
is a sequence of finitely generated sub-$\sigma$-fields.
It is therefore clear from a concrete description of
the Radon-Nikodym derivative in terms of a limit of 
$\cF_k$-measurable ratios (see, for example,
Section~III-1 of \cite{MR0402915})
that we may suppose that the map 
$(t,m) \to \phi_t(m)$ is Borel measurable.  

By Fubini's theorem, 
\begin{equation*}
\int_A \phi_t(m) \, \nu(dm) 
= \int_A \phi_0(m) \, \nu(dm) 
+ t \int_A \, \nu(dm) 
- \int_A \left[\int_0^t F_{\rho_s} (m) \phi_s(m) \, ds \right]\, \nu(dm)
\end{equation*}
for any Borel set $A \subseteq \cM$.
Thus, for each $t \ge 0$ we have for $\nu$-a.e. $m \in \cM$ that 
\begin{equation}
\label{ptwise}
\phi_t(m) 
= \phi_0(m) + t - \int_0^t F_{\rho_s}(m)\phi_s(m) \, ds
=: \psi_t(m),
\end{equation}
by the uniqueness of the Radon-Nikodym derivative.
A fortiori, (\ref{ptwise}) holds for 
$\lambda \otimes \nu$-a.e. pair $(t,m) \in \bR_+ \times \cM$ 
(recall that $\lambda$ is Lebesgue measure on $\bR_+$). 
Thus, by Fubini's theorem, it follows that, 
for $\nu$-a.e. $m \in \cM$,  the equation (\ref{ptwise})
holds for $\lambda$-a.e. $t \ge 0$.
Also, $\rho_t$ is the measure 
$\breve \psi_t(dm): = \psi_t(m)\, \nu(dm)$ 
for all $t \ge 0$, and hence
$F_{\rho_t}(m) = F_{\breve \psi_t}(m)$ for all $t \ge 0$.
Thus, for $\nu$-a.e. $m \in \cM$ and for all $t \ge 0$, 
\begin{equation*}
\int_0^t F_{\breve \psi_s}(m)\psi_s(m) \, ds = \int_0^t F_{\rho_s}(m) \phi_s(m) \, ds.
\end{equation*}

By definition, $\psi_0 = \phi_0$, and so, for $\nu$-a.e. $m \in \cM$ and for all $t \ge 0$, 
\[
\label{inteq}
\psi_t(m) 
= \psi_0(m) + t - \int_0^t F_{\breve \psi_s}(m) \psi_s(m) \, ds.
\]
Therefore, 
for $\nu$-a.e. $m \in \cM$ the function $t \mapsto \psi_t(m)$
is differentiable with
\[
\dot{\psi}_t(m) 
= 1 - F_{\breve \psi_t}(m) \psi_t(m) 
= 1 - G \psi_t (m),
\]
and setting $x_t = \psi_t$ gives the result.
\end{proof}

\section{Equilibria and Stability}
\label{S:equilibrium}

We can now state our main theorem.
\begin{Theo}
\label{stability}
Suppose that $E:= \inf \{a_1(m) \; : \; m \in \cM\}>0$.
Then the system  (\ref{sys}) has a 
unique equilibrium point in the space $L^\infty_+(\cM,\nu)$.  Furthermore, this equilibrium is globally stable, in the sense that all solutions to (\ref{sys}) converge to it in the $L^\infty(\cM,\nu)$ norm as $t$ goes to infinity.
\end{Theo}

The proof will be accomplished in four parts.  First, we derive some upper and lower bounds on solutions to (\ref{sys}).  Then, we
establish existence of at least one equilibrium point by
a fixed point argument and show
uniqueness using a generalization of a 
method of Craciun and Feinberg \cite{MR2177713}.  Finally, we
obtain global stability of the equilibrium via the
construction of a Lyapunov function.  Since we are dealing with an infinite dimensional dynamical system, all of these steps rely on some technical lemmas concerning the continuity of 
certain operators on $L^\infty_+(\cM,\nu)$, and
we defer these to Section~\ref{S:weak_cty}.

\subsection{Estimates on the solution}  

Let $x_t$ be a solution to the equation (\ref{sys}).
Since $G x_t (m) \geq a_1(m) x_t(m)$, we see 
for $\nu$-a.e. $m \in \cM$ and all $t \ge 0$ that
$\dot{x}_t(m) \leq 1 - a_1(m)x_t(m)$. Hence,
\begin{equation}
\label{first_upperbd}
x_t(m) \le \displaystyle{\frac{1-(1-a_1(m)x_0(m)) e^{-a_1(m)t}}{a_1(m)}}
\end{equation}
for $\nu$-a.e. $m \in \cM$ and all $t \ge 0$.  As $t \to \infty$ the
right-hand side of (\ref{first_upperbd})
either increases to the value $1/a_1(m)$, in the case that 
$x_0(m) < 1/a_1(m)$, or otherwise decreases to the same value.  
In either case, we have for $\nu-$a.e. $m \in \cM$
that $x_t(m) \leq \max \{x_0(m),1/a_1(m)\}$ for all $t \ge 0$.  Thus, 
\begin{equation}
\label{upperbd}
\sup_t \|x_t\|_\infty \le K, 
\end{equation}
where $K = \|x_0\|_\infty \vee A$ with $A:= E^{-1}$.

Similarly, since for all $t \ge 0$ we have
$x_t(m) \le K$ for $\nu$-a.e. $m \in \cM$, it follows that 
\begin{equation}
\label{Gbound}
G x_t(m) 
\le 
\left[
\sum_{n=1}^N n \int_{\cM^{(n-1)}} a_n(\mathbf{m},m) 
K^{(n-1)}  \nu ^{\otimes (n-1)}(d\mathbf{m}) 
\right]
x_t(m)
\le C x_t(m), 
\end{equation}
for some constant $C$. Therefore, 
\begin{equation*}
\begin{split}
x_t(m) 
& \ge \frac{1-(1-C x_0(m))e^{-Ct}}{C} \\
& \ge \frac{1-e^{-Ct}}{C} + x_0(m)e^{-Ct}
\geq \frac{1-e^{-Ct}}{C}. \\
\end{split}
\end{equation*}
Hence,
\begin{equation}
\label{lowerbd}
\liminf_{t \to \infty} \ess \inf_m x_t(m) \geq 1/C > 0.
\end{equation}

\subsection{Existence of equilibria}
\label{SS:exist}

Observe that $x \in L_+^\infty(\cM, \nu)$ is an 
equilibrium point of (\ref{sys}) if and only if $G x(m)=1$ for $\nu$-a.e. $m \in \cM$, that is, if and only if $x$ is a fixed point of the map 
$\Gamma:L^\infty_+(\cM, \nu) \to L^\infty_+(\cM, \nu)$ given by
\begin{equation}
\label{def_Gamma}
\Gamma y(m) := 
\left[
\sum_{n=1}^N n \int_{\cM^{(n-1)}} a_n(\mathbf{m},m) y(m_1) \cdots y(m_{n-1}) \, \nu^{\otimes (n-1)} (d\mathbf{m})
\right]^{-1}.
\end{equation}

Note for any $y \in L^\infty_+(\cM, \nu)$ that
\[
a_1(m) 
\le \sum_{n=1}^N n \int_{\cM^{(n-1)}} a_n(\mathbf{m},m) y(m_1) \cdots y(m_{n-1}) \, \nu^{\otimes (n-1)} (d\mathbf{m}).
\]
Hence,
\begin{equation}
\label{always_le_A}
\Gamma y (m) \le  A, \quad \text{for $\nu$-a.e. $m \in \cM$.}
\end{equation}
Thus $\Gamma$ does indeed map
$L^\infty_+(\cM, \nu)$ into itself.
Moreover, if  $y(m) \le A$ for $\nu$-a.e. $m \in \cM$,
then
\begin{equation*}
\begin{split}
&\sum_{n=1}^N n \int_{\cM^{(n-1)}} a_n(\mathbf{m},m) y(m_1) \cdots y(m_{n-1}) \, \nu^{\otimes (n-1)} (d\mathbf{m}) \\
& \quad \le \sum_{n=1}^N n \int_{\cM^{(n-1)}} a_n(\mathbf{m},m) A^{(n-1)} \, \nu^{\otimes (n-1)} (d\mathbf{m}) =: B^{-1}, \\
\end{split}
\end{equation*}
so that 
\begin{equation}
\label{always_ge_B}
B \le  \Gamma y(m), \quad \text{for $\nu$-a.e. $m \in \cM$.}
\end{equation}

Therefore, $\Gamma$ maps the convex set 
\begin{equation}
\label{def_R}
R := \{y \in L^\infty_+(\cM,\nu) \; : \; B \le y(m) \le A \}
\end{equation} 
into itself.  
Recall that the Banach space $L^\infty(\cM,\nu)$
is the dual of the Banach space $L^1(\cM,\nu)$
under the pairing 
$\langle x,f \rangle :=  \int_\cM x(m) f(m) \, \nu(dm)$ for
$x \in L^\infty(\cM,\nu)$ and $f \in L^1(\cM,\nu)$. Recall also that
the weak* topology on $L^\infty(\cM,\nu)$ is the weakest
topology that makes each of the maps 
$x \mapsto \langle x, f \rangle$, $f \in L^1(\cM,\nu)$,
continuous.  A consequence of the 
Banach-Alaoglu Theorem is that that any weak*-closed and norm-bounded subset of 
$L^\infty(\cM,\nu)$ is weak*-compact
(see Corollary 3 of Section V.4.2 of \cite{MR1009162}),
and so the set
$R$ is weak*-compact.  
The map $\Gamma$ is weak*-continuous on the convex
weak*-compact set $R$
by Lemma~\ref{Fcont} below. 
An infinite-dimensional extension of the 
Brouwer Fixed Point Theorem, the Schauder-Tychonoff Theorem 
(see Theorem 5 of Section V.10.5 in \cite{MR1009162}),
guarantees that $\Gamma$ has a fixed point in $R$.  
Consequently, the system (\ref{sys}) has at least one equilibrium point.

\begin{Rem}
\label{R:equilibrium_in_R}
Note from the argument above that if
$x \in L^\infty_+(\cM,\nu)$
is any equilibrium point of (\ref{sys}), then
$x(m) = \Gamma x(m) \le A$
for $\nu$-a.e. $m \in \cM$
by (\ref{always_le_A}), 
and so $x \in R$ by (\ref{always_ge_B}).
\end{Rem}

\subsection{Uniqueness of the equilibrium}
\label{SS:unique}

We claim that there is only one solution in $L^\infty_+(\cM,\nu)$
to the equilibrium equation $G x(m)=1$, $\nu$-a.e. $m \in \cM$.  
Our argument follows that used
in Theorem 3.1 of  \cite{MR2177713} to establish
a criterion for the uniqueness of equilibria in the finite dimensional {\em mass-action kinetics}
systems of differential equations that arise in the
study of {\em continuous flow stirred tank reactors}.  
We refer
the reader to \cite{MR2177713} for a discussion of the history of related results.  A recent paper on the use of
algebraic geometry in analyzing instances of 
such  systems is
\cite{CrDiShSt07}, which also contains several references
to other applications of this general class of dynamical systems.

By Remark~\ref{R:equilibrium_in_R}, 
it suffices to show that there is only one solution in 
the set $R$.  
Assume otherwise, that is, that there are two solutions $x \in R$ and $y \in R$ with $x \neq y$.  In particular, we have that $G x(m)=G y(m)$ for $\nu$-a.e. $m \in \cM$, and so
\begin{equation*}
\begin{split}
& \sum_{n=1}^N n \int_{\cM^{(n-1)}} a_n(\mathbf{m},m) x(m_1) \cdots x(m_{n-1}) \,  x(m)\, \nu ^{\otimes (n-1)}(d\mathbf{m})\\
& \quad = \sum_{n=1}^N n \int_{\cM^{(n-1)}} a_n(\mathbf{m},m) y(m_1) \cdots y(m_{n-1})\, y(m) \,  \nu ^{\otimes (n-1)}(d\mathbf{m}).\\
\end{split}
\end{equation*}
Therefore, 
\begin{equation*}
\begin{split}
& \sum_{n=1}^N n \int_{\cM^{(n-1)}} a_n(\mathbf{m},m) x(m_1) \cdots x(m_{n-1})x(m) \\
& \quad \times \left(\frac{y(m_1) \cdots y(m_{n-1})y(m)}{x(m_1) \cdots x(m_{n-1})x(m)}-1 \right)
\, \nu ^{\otimes (n-1)}(d\mathbf{m})=0 \\
\end{split}
\end{equation*}
for $\nu$-a.e. $m \in \cM$.  Setting 
$K_n(\mathbf{m},m) :=a_n(\mathbf{m},m) x(m_1) \cdots x(m_{n-1})x(m)$ and 
$\delta(m) := \log \left(y(m)/x(m)\right)$,  we obtain
\begin{equation*}
\sum_{n=1}^N n \int_{\cM^{(n-1)}} 
K_n(\mathbf{m},m)  \left(e^{\delta(m_1)+ \cdots +\delta(m_{n-1})+\delta(m)} - 1 \right)
\, \nu ^{\otimes (n-1)}(d\mathbf{m})=0.
\end{equation*}
Observe that $\delta(m)$ is bounded since $x$ and $y$ are in $R$.
Thus, putting 
\[
\eta_n(\mathbf{m},m) := 
K_n(\mathbf{m},m) \frac{e^{\delta(m_1)+ \cdots +\delta(m_{n-1})+\delta(m)} - 1}{\delta(m_1)+ \cdots +\delta(m_{n-1})+\delta(m)}, 
\]
we get 
\begin{equation}
\label{delta}
\sum_{n=1}^N n \int_{\cM^{(n-1)}} \eta_n(\mathbf{m},m)  \left(\delta(m_1)+ \cdots +\delta(m_{n-1})+\delta(m) \right)\nu ^{\otimes (n-1)}(d\mathbf{m})=0
\end{equation}
for $\nu$-a.e. $m \in \cM$.  Note that the function
$\eta_n$ is non-negative, since $\delta(m_1)+ \cdots +\delta(m_{n-1})+\delta(m)$ and $e^{\delta(m_1)+ \cdots +\delta(m_{n-1})+\delta(m)} - 1$ have the same sign.  Also, $\eta_n$ is permutation invariant and takes the
value $0$ whenever two of the coordinates of $\mathbf{m}$ are
equal.

Integrating the left-hand side of (\ref{delta}) 
against the function $\delta$ gives
\begin{equation*}
\begin{split}
0
& =
\int_\cM\sum_{n=1}^N n \int_{\cM^{(n-1)}} \eta_n(\mathbf{m},m) \\
& \quad \times \left(\delta(m_1)+ \cdots +\delta(m_{n-1})+\delta(m) \right)\nu ^{\otimes (n-1)}(d\mathbf{m})  \delta(m) \, \nu(dm) \\
& =
\sum_{n=1}^N n\int_{\cM^n} \eta_n(m_1,\ldots,m_n)(\delta(m_1)+ \cdots +\delta(m_{n-1})+\delta(m_n)) \\
& \quad \times \delta(m_n) \nu ^{\otimes n}(dm_1,\ldots,dm_n) \\
& = 
\sum_{n=1}^N \sum_{k=1}^n \int_{\cM^n} \eta_n(m_1,\ldots,m_n)(\delta(m_1)+ \cdots +\delta(m_n)) \\
& \quad \times \delta(m_k) 
\, \nu ^{\otimes n}(dm_1,\ldots,dm_n), \\
\end{split}
\end{equation*}
by the symmetry of $\eta_n$ together with the change of variables 
\[
(m_1,\ldots,m_k,\ldots,m_n) 
\leftrightarrow (m_1,\ldots,m_n,\ldots,m_k)
\] 
once for each $k$.

Therefore, 
\begin{equation*}
0=\sum_{n=1}^N \int_{\cM^n} \eta_n(m_1,\ldots,m_n) \left[\delta(m_1)+ \cdots +\delta(m_n) \right]^2 \, 
\nu ^{\otimes n}(dm_1,\ldots,dm_n).
\end{equation*}
In particular,
$
0=\int_\cM \eta_1(m)\delta(m)^2 \, \nu(dm)
$.
Since $a_1(m) \ge E > 0$ for all $m \in \cM$, 
it follows that $\eta_1(m) > 0$ for all  $m \in \cM$, and hence $\delta(m)=0$ for $\nu$-a.e. $m \in \cM$.   This contradicts our
assumption that $x \ne y$, and so the equilibrium point is unique.

\subsection{Stability}

Let  $x_t$ be a solution of (\ref{sys}).
Write $x^* \in R \subset L^\infty_+(\cM,\nu)$ 
for the unique equilibrium point guaranteed
by Subsection~\ref{SS:exist} and Subsection~\ref{SS:unique}.
We will show that 
$\lim_{t \rightarrow \infty} \|x_t - x^*\|_\infty = 0$
using a variant of {\em Lyapunov's second method}.  

A recent review of Lyapunov function methods for studying
the asymptotic behavior of finite dimensional
nonlinear systems is \cite{MR2104692}.  Much of the
finite dimensional theory considers Lyapunov functions that
are continuously differentiable.  General stability results
for finite dimensional
systems that do not assume this degree of smoothness
are discussed in \cite{MR1754911}, where there is
a number of references indicating why weaker
assumptions are natural in several applications.  Infinite dimensional
systems are considered in \cite{MR513814}, primarily in the
context of partial differential equations. Discontinuous
Lyapunov functionals appear there because it is natural to
work with a weak* topology in order to make norm-bounded sets
compact, but functionals that are continuous in the norm
topology may cease to be continuous in the weak* topology.
This is exactly the situation that confronts us.  
Because we haven't found a result in the literature that
applies directly to establish global stability in our setting,
we provide the details of the relatively routine argument.

Define the function $V:L^\infty_+(\cM,\nu) \to \bR \cup \{+ \infty\}$  by 
\begin{equation}
\begin{split}
V(x) & := - \int_\cM \log(x(m))\, \nu(dm)  \\
& \quad + \sum_{n=1}^N \int_{\cM^n} a_n(\mathbf{m}) x(m_1) \cdots x(m_n) \, \nu ^{\otimes n} (d \mathbf{m}).
\end{split}
\end{equation}
Observe that $V$ is, indeed, $\bR \cup \{+ \infty\}$-valued and even bounded below, because
\begin{equation}
\label{lowerbd_V}
\begin{split}
V(x)& = - \int_\cM \log(x(m))\, \nu(dm) + \int_\cM a_1(m)x(m) \, \nu(dm) \\ 
& \quad + \sum_{n=2}^N \int_{\cM^n} a_n(\mathbf{m}) x(m_1) \cdots x(m_n) \, \nu ^{\otimes n} (d \mathbf{m}) \\
&  \geq \int_\cM \left[a_1(m)x(m) - \log(x(m))\right] \, \nu(dm) \\
&  \geq \int_\cM \left[1+\log(a_1(m))\right]\, \nu(dm) \\
&  \ge (1+\log(E)) \times \nu(\cM)  > -\infty,\\
\end{split}
\end{equation}
since $a u-\log(u) \ge 1+\log(a)$ for all $u > 0$.

Set $V_t := V(x(t))$.  We have 
\begin{equation}
\label{vdotdef}
\begin{split}
&\frac{d}{d t} V_t 
 = 
\lim_{h \to 0} \frac{1}{h} \left[ -\int_\cM \log(x_{t+h}(m)) \, \nu(dm) + \int_\cM \log(x_t(m)) \, \nu(dm) \right] \\
& \quad + 
\lim_{h \to 0} \frac{1}{h} \left[ \int_\cM a_1(m)x_{t+h}(m) \, \nu(dm) - \int_\cM a_1(m)x_t(m)\, \nu(dm)\right] \\
& \quad + 
\lim_{h \to 0} \frac{1}{h} \sum_{n=2}^N \int_{\cM^n} a_n(\mathbf{m}) \\
& \qquad \times \left[x_{t+h}(m_1) \cdots x_{t+h}(m_n)-x_t(m_1) \cdots x_t(m_n)\right] \, \nu^{\otimes n}(d\mathbf{m}). \\
\end{split}
\end{equation}
From (\ref{upperbd}) and (\ref{lowerbd}) there exist constants
$C>0$, $K>0$, and
$T \ge 0$ such that
\begin{equation}
\frac{1}{2C} \le \ess \inf_m x_t(m) \le \ess \sup_m x_t(m) \le K, \quad t \ge T.
\end{equation}  
Hence, by the Dominated Convergence Theorem, we may interchange the limit and integrals in (\ref{vdotdef}) to get
\begin{equation}
\label{vdot}
\begin{split}
& \frac{d}{d t} V_t
 = 
-\int_\cM \frac{\dot{x}_t(m)}{x_t(m)} \, \nu(dm) + \int_\cM a_1(m) \dot{x}_t(m) \, \nu(dm) \\
& \qquad + 
\sum_{n=2}^N n \int_{\cM^n} a_n(\mathbf{m}) x_t(m_1) \cdots x_t(m_{n-1}) \dot{x}_t(m_n) \, \nu^{\otimes n} (d\mathbf{m}) \\
& \quad =
\int_\cM (-\dot{x}_t(m)) \biggl[\frac{1}{x_t(m)} - a_1(m) \\
& \qquad - \sum_{n=2}^N n \int_{\cM^{(n-1)}} a_n(\mathbf{m}) x_t(m_1) \cdots x_t(m_{n-1}) \, \nu^{\otimes (n-1)} (d\mathbf{m}) \biggr] \, \nu(dm) \\
& \quad = 
- \int_\cM \dot{x}_t(m)\frac{1}{x_t(m)} \dot{x}_t(m) \, \nu(dm) \\
& \quad \le 0 \\
\end{split}
\end{equation}
for all $t \ge T$.

Define $\dot{V}: L^\infty_+(\cM,\nu) \rightarrow
\bR_- \cup \{-\infty\}$ by
\begin{equation}
\label{def_dotV}
\dot{V}(x) := -\int_\cM (1-G x(m))^2 \frac{1}{x(m)} \, \nu(dm).
\end{equation}
From (\ref{vdot}) and (\ref{sys}), we obtain
$\frac{d}{d t} V_t = \dot{V}(x_t)$ for  $t \ge T$.

By Lemma~\ref{vdotsemi}, the function
$\dot{V}$ is upper semicontinuous in the weak* topology on the domain 
$\{x \in L_+^\infty(\cM,\nu) : \frac{1}{2C} \le \ess \inf_m x(m) \le \ess \sup_m x(m) \le K\}$.  
Also, if $x$ is in this domain, then $\dot{V}(x) = 0$ 
if and only if $G x(m) = 1$ for $\nu-$a.e. $m \in \cM$, that is, if and only if $x=x^*$.   
It follows that for any $\epsilon >0$ 
the supremum of $\dot{V}$ on the weak* compact set 
\begin{equation*}
\{x \in L_+^\infty(\cM,\nu) : 
\frac{1}{2C} \le \ess \inf_m x(m) \le \ess \sup_m x(m) \le K, \, 
\|x-x^*\|_\infty \geq \epsilon\}
\end{equation*}
is strictly less than 0.

Assume that $\|x_t - x^*\|_\infty$ does not converge to $0$
as $t \rightarrow \infty$.  Then there exists $\delta>0$ and a sequence $t_k \to \infty$ with $t_k \ge T$
for all $k$ such that $\|x_{t_k} - x^*\|_\infty > \delta$ for all $k$.  Now, since $\dot{x}_t$ is uniformly bounded
as a consequence of (\ref{upperbd}),  (\ref{Gbound}), and
(\ref{lowerbd}),
there is a number $\gamma >0$ independent of $k$ such that $\|x_t-x_{t_k}\|_\infty < \delta/2$ for $t_k \le t \le t_k + \gamma$, so that $\|x_t - x^*\|_\infty > \delta/2$ for $t_k \le t \le t_k + \gamma$.  It follows that there is a constant $\lambda < 0$ independent of $k$ such that 
\begin{equation}
\label{vdotest}
\frac{d}{dt} V_t = \dot{V}(x_t) \le \lambda < 0
\end{equation}
for $t_k \le t \le t_k + \gamma$.
Because $V_t$ is nondecreasing for
$t \ge T$ by (\ref{vdot}), the inequality (\ref{vdotest}) contradicts the conclusion from
(\ref{lowerbd_V}) that $\inf_t V_t > -\infty$.  Thus
$\|x_t - x^*\|_\infty$ must converge to $0$
as $t \rightarrow \infty$.

\begin{Rem}
Put $\bar \nu := c^{-1} \nu$, where $c:= \nu(\cM)$, and 
$
\dot{y}_t := \frac{d \rho_{ct}}{d \bar \nu} = c x_{c^{-1} t}
$.
Then, 
$
\dot{y}_t(m) = \dot{x}_{c^{-1} t}(m) = 1 - \bar G y_t(m)
$,
where $\bar G$ is defined in the same way as $G$, except that
$\nu$ is replaced by $\bar \nu$.  Therefore, by a change of time
units, we may suppose without loss of generality that $\nu$
is a probability measure.  In that case, the Lyapunov function $V$
may be written as
\[
V(x) = D(\nu \, \| \, \bar x \cdot \nu) 
- \log\left(\int x(m) \, \nu(dm)\right) 
+ \bE \left[S(X^{x \cdot \nu})\right],
\]
where $D(\cdot \, \| \, \cdot)$ denotes the relative entropy or
Kullback-Leibler distance between two probability measures,
$\bar x := (\int x(m) \, \nu(dm))^{-1} x$ is $x$ renormalized to
be a probability density with respect to the measure $\nu$, 
$\bar x \cdot \nu$ is the probability measure with density $\bar x$
with respect to $\nu$, $x \cdot \nu$ is the finite measure with density $ x$
with respect to $\nu$, and $X^{x \cdot \nu}$ is a Poisson random measure
with intensity $x \cdot \nu$.  Therefore, our system (\ref{model}) evolves
in such a way that it tries to  decrease the sum of
the expected selective cost of an individual chosen at random from the population,
 $\bE [S(X^{\rho_t})]$,
and the Kullback-Leibler distance between the probability measures
$\nu$ and $\rho_t(\cM)^{-1} \rho_t$,
while not allowing the expected total number of ancestral
mutation events present in the  of a randomly chosen individual,
$\rho_t(\cM) = \bE[X^{\rho_t}(\cM)]$, to be too small.
\end{Rem}

We note that the sum of a Kullback-Leibler distance
and the logarithm of expected selective cost
appears as a Lyapunov function for
the continuous-time,
single-locus, $n$-allele, mutation-selection model
of \cite{MR821683} in which the mutation rate
from allele $i$ to allele $j$ only depends on
the target allele $j$.  This Lyapunov
function is used there to prove a global stability result.
Also, a Kullback-Leibler distance is a Lyapunov function
in a neighborhood of the stable state of 
the frequency-dependent selection model of 
\cite{MR1130789} that builds on the 
evolutionary game theory work
of \cite{MR658973}, 
where a similar quantity also appears.

\section{Some technical lemmas}
\label{S:weak_cty}

We collect together in this section some results
used in the proof of Theorem~\ref{stability}.

The following result is elementary, but we include it for
completeness.

\begin{Lemma}
\label{cty_multiple_int}
Suppose that $f \in L^1(\cM^n, \nu^{\otimes n})$ for
some positive integer $n$.  The function
\begin{equation*}
x \mapsto \int_{\cM^n} 
x(m_1) \cdots x(m_n) f(m_1, \ldots, m_n)
\, \nu^{\otimes n}(d\mathbf{m})
\end{equation*}
is weak*-continuous on any norm bounded subset
of $L_+^\infty(\cM, \nu)$.
\end{Lemma}

\begin{proof}
This is obvious for functions of the form 
\[
f(m_1, \ldots, m_n) = f_1(m_1) \ldots f_n(m_n),
\] where
$f_1, \ldots, f_n \in L^1(\cM,\nu)$, and the result
for general $f$ follows from the fact that finite linear
combinations of such functions are dense in 
$L^1(\cM^n, \nu^{\otimes n})$.
\end{proof}

Recall the definition of the map
$\Gamma: L^\infty_+(\cM,\nu) \to L^\infty_+(\cM,\nu)$ 
and the set $R \subset L_+^\infty(\cM,\nu)$ 
from (\ref{def_Gamma}) and
(\ref{def_R}), respectively. Recall also that
$\Gamma$ maps $R$ into itself.

\begin{Lemma}
\label{Fcont}
The map $\Gamma:R \rightarrow R$ is weak*-continuous.
\end{Lemma}

\begin{proof}
Suppose that 
$x_k$ is a sequence in $R$ such that $x_k$ converges to 
$x \in R$ in the weak* topology as $k \to \infty$.  
Fix a test function $f \in L^1(\cM,\nu)$.
Then 
\begin{equation*}
\begin{split}
& \left |
\int_\cM  \Gamma x_k(m)f(m) \, \nu(dm)
- \int_\cM \Gamma x(m)f(m) \, \nu(dm) \right| \\
& \quad = \Bigl| \sum_{n=2}^N n 
\int_{\cM} f(m)\frac{1}{H(m)} \\ 
& \qquad \times
\int_{\cM^{(n-1)}}
 a_n(\mathbf{m},m)
 \left[ 
 x_k(m_1)\cdots x_k(m_{n-1}) - x(m_1)\cdots x(m_{n-1})
 \right]
 \, \nu^{\otimes (n-1)}(d\mathbf{m}) \\
& \qquad \times \, \nu(dm) \Bigr|,  \\
\end{split}
\end{equation*}
where 
\begin{equation*}
\begin{split}
H(m) 
& := \left(\sum_{n=1}^N n \int_{\cM^{(n-1)}} a_n(\mathbf{m},m) x_k(m_1)\ldots x_k(m_{n-1}) \, \nu^{\otimes (n-1)} (d\mathbf{m})\right) \\
& \times \left(\sum_{n=1}^N n \int_{\cM^n_m} a_n(\mathbf{m}) x(m_1)\ldots x(m_{n-1}) \, \nu^{\otimes (n-1)} (d\mathbf{m}) \right) \ge E^2. \\
\end{split}
\end{equation*}

Therefore,
\begin{equation*}
\begin{split}
& \left| 
\int_\cM \Gamma x_k(m)f(m) \, \nu(dm)
- \int_\cM \Gamma x(m)f(m) \, \nu(dm) 
\right| \\
& \quad \le \frac{1}{E^2} \sum_{n=2}^N \int_{\cM} |f(m)| \\
& \qquad \times
\left|
\int_{\cM^{(n-1)}} a_n(\mathbf{m},m)
\left[
x_k(m_1)\cdots x_k(m_{n-1}) - x(m_1) \cdots x(m_{n-1})
\right]
\, \nu^{\otimes (n-1)}(d\mathbf{m})
\right| \\
& \qquad \times 
\, \nu(dm).\\
\end{split}
\end{equation*}

Each integral 
\begin{equation*}
\int_{\cM^{(n-1)}} a_n(\mathbf{m},m)
\left[
x_k(m_1)\cdots x_k(m_{n-1})-x(m_1) \cdots x(m_{n-1})
\right]
\, \nu^{\otimes (n-1)}(d\mathbf{m})
\end{equation*}
belongs to $L^\infty(\cM,\nu)$ with norm that is bounded
in $k$.  Moreover, each
such integral converges to zero as $k$ goes to infinity
by Lemma~\ref{cty_multiple_int}.  It follows from the
Dominated Convergence Theorem that
\begin{equation*}
\lim_{k \rightarrow \infty}
\left| 
\int_\cM \Gamma x_k(m)f(m) \, \nu(dm)
- \int_\cM \Gamma x(m)f(m) \, \nu(dm) 
\right|
= 0,
\end{equation*}
as required.
\end{proof}

Recall the definition of the function
$\dot{V}: L^\infty_+(\cM,\nu) \rightarrow \bR$ 
from (\ref{def_dotV}).

\begin{Lemma}
\label{vdotsemi}
For any constants $0 < a \le b < \infty$, 
the function $\dot{V}$ restricted to the set
\begin{equation*}
\{x \in L^\infty_+(\cM,\nu) : a \le \ess \inf_m x(m) \le
\ess \sup_m x(m) \le b\}
\end{equation*}
is upper semicontinuous in the weak* topology.
\end{Lemma}

\begin{proof} 
Fix a domain $D \subset L_+^\infty(\cM,\nu)$ of the type 
considered in the statement of the lemma.

Observe that
\begin{equation*}
\dot{V}(x) = -\int_\cM \frac{1}{x(m)}\, \nu(dm) + H(x),
\end{equation*}
where $H: D \rightarrow \bR$ 
is a sum of functions, each of the form 
\begin{equation*}
x \mapsto \int g(m_1,\ldots,m_k) x(m_1) \cdots x(m_k) \, \nu(dm_1) \cdots \, \nu(dm_k)
\end{equation*}
for some function $g \in L^\infty(\cM^k, \nu^{\otimes k})$.
It therefore suffices by Lemma~\ref{cty_multiple_int} to show that the function 
\begin{equation}
x \mapsto -\int_\cM \frac{1}{x(m)}\, \nu(dm)
\end{equation}
is weak*-upper semicontinuous 
on $D$  

Suppose without loss of generality that 
$\nu$ is a probability measure.    
Let $\Pi_n = (A_{n_1},\ldots,A_{n_{p(n)}})$ 
be a sequence of partitions of $\cM$ such that the $\sigma$-fields generated by the successive
partitions form a filtration $\cF_n$ with
the property that
$\sigma(\bigcup_n \cF_n) = \cB(\cM)$, the Borel $\sigma$-field on $\cM$.  Consider $x \in D$  
as a random variable on the probability space
$(\cM, \cB(\cM), \nu)$, and let 
$x^{(n)} :=\bE[x \, | \,  \cF_n]$
be the conditional expectation of $x$ given $\cF_n$.  Note that $x^{(n)}$ is a non-negative martingale
with $a \le x^{(n)}(m) \le b$ for $\nu$-a.e. $m \in \cM$.  
For $m \in A_{n_q}$ we have
\begin{equation*}
x^{(n)}(m) = \frac{1}{\nu(A_{n_q})}\int_{A_{n_q}} x(m)\, \nu(dm),
\end{equation*}
with the convention that $0/0 = 0$.
The function taking $x$ to each such integral is weak*-continuous.  
As a result, the function 
\begin{equation*}
x \mapsto - \int \frac{1}{x^{(n)}(m)} \, \nu(dm) 
= - \sum_q \nu(A_{n_q}) 
\left[
\frac{1}{\nu(A_{n_q})} \int_{A_{n_q}} x(m)\, \nu(dm) 
\right]^{-1}
\end{equation*}
is also weak*-continuous.  

By Jensen's inequality for conditional expectation, 
the sequence of random variables $-1/x^{(n)}$ is a non-positive supermartingale, and so 
the sequence of expectations
\[
-\int_\cM [x^{(n)}(m)]^{-1}\, \nu(dm) = \mathbb{E}[-1/x^{(n)}]
\] 
is non-increasing.  
By the Martingale Convergence Theorem 
(see, for example, Theorem~IV-1-2 of \cite{MR0402915}),
the sequence $x^{(n)}(m)$ converges to $x(m)$  
for $\nu$-a.e. $m \in \cM$.
 Therefore, $-\int_\cM [x^{(n)}(m)]^{-1} \, \nu(dm)$ decreases to $-\int_\cM x(m)^{-1} \, \nu(dm)$ by the
Dominated Convergence Theorem.  

We can thus write the function 
$x \mapsto -\int_\cM x(m)^{-1} \, \nu(dm)$ 
as the infimum of the family of functions 
$x \mapsto -\int_\cM [x^{(n)}(m)]^{-1} \, \nu(dm)$, 
each of which is weak*-continuous.  Consequently, the function
$x \mapsto -\int_\cM x(m)^{-1} \, \nu(dm)$ is weak*-upper semicontinuous.
\end{proof}

\bigskip 
\noindent{\bf Acknowledgment:}  Parts of this paper
were written while the second author was a Visiting Fellow
at the Mathematical Sciences Institute of the
Australian National University.  He
thanks the members of the Institute for their
hospitality.  The authors also thank 
Reinhard B\"urger,
Bernd Sturmfels, David Steinsaltz and
Ken Wachter for numerous helpful suggestions.

\providecommand{\bysame}{\leavevmode\hbox to3em{\hrulefill}\thinspace}
\providecommand{\MR}{\relax\ifhmode\unskip\space\fi MR }
% \MRhref is called by the amsart/book/proc definition of \MR.
\providecommand{\MRhref}[2]{%
  \href{http://www.ams.org/mathscinet-getitem?mr=#1}{#2}
}
\providecommand{\href}[2]{#2}


\begin{thebibliography}{CDSS07}

\bibitem[Aki82]{MR658973}
Ethan Akin, \emph{Exponential families and game dynamics}, Canad. J. Math.
  \textbf{34} (1982), no.~2, 374--405. \MR{MR658973 (83m:92052)}

\bibitem[Baa01]{MR1842838}
Ellen Baake, \emph{Mutation and recombination with tight linkage}, J. Math.
  Biol. \textbf{42} (2001), no.~5, 455--488. \MR{MR1842838 (2002d:92008)}

\bibitem[Baa05]{MR2191729}
Michael Baake, \emph{Recombination semigroups on measure spaces}, Monatsh.
  Math. \textbf{146} (2005), no.~4, 267--278. \MR{MR2191729 (2006k:92055)}

\bibitem[Baa07]{MR2297255}
\bysame, \emph{Addendum to: ``{R}ecombination semigroups on measure spaces''
  [{M}onatsh. {M}ath. {\bf 146} (2005), no. 4, 267--278; mr2191729]}, Monatsh.
  Math. \textbf{150} (2007), no.~1, 83--84. \MR{MR2297255}

\bibitem[BB96]{MR1372337}
Reinhard B{\"u}rger and Immanuel~M. Bomze, \emph{Stationary distributions under
  mutation-selection balance: structure and properties}, Adv. in Appl. Probab.
  \textbf{28} (1996), no.~1, 227--251. \MR{MR1372337 (97h:92008)}

\bibitem[BB03]{MR1952324}
Michael Baake and Ellen Baake, \emph{An exactly solved model for mutation,
  recombination and selection}, Canad. J. Math. \textbf{55} (2003), no.~1,
  3--41. \MR{MR1952324 (2004a:92015)}

\bibitem[Bom91]{MR1130789}
Immanuel~M. Bomze, \emph{Cross entropy minimization in uninvadable states of
  complex populations}, J. Math. Biol. \textbf{30} (1991), no.~1, 73--87.
  \MR{MR1130789 (92j:92012)}

\bibitem[BT91]{BaTa91}
N.~H. Barton and Michael Turelli, \emph{Natural and sexual selection on many
  loci}, Genetics \textbf{127} (1991), 229–--255.

\bibitem[B{\"u}r98]{Bur98}
Reinhard B{\"u}rger, \emph{Mathematical properties of mutation selection
  models}, Genetica \textbf{102/103} (1998), 279–--298.

\bibitem[B{\"u}r00]{MR1885085}
\bysame, \emph{The mathematical theory of selection, recombination, and
  mutation}, Wiley Series in Mathematical and Computational Biology, John Wiley
  \& Sons Ltd., Chichester, 2000. \MR{MR1885085 (2002m:92002)}

\bibitem[BW01]{BaWa01}
Ellen Baake and Holger Wagner, \emph{Mutation{–-}selection models solved
  exactly with methods of statistical mechanics}, Genet. Res., Camb.
  \textbf{78} (2001), 93–--117.

\bibitem[CDSS07]{CrDiShSt07}
Gheorghe Craciun, Alicia Dickenstein, Anne Shiu, and Bernd Sturmfels,
  \emph{Toric dynamical systems}, 2007, Preprint, available at {\tt
  http://arxiv.org/abs/0708.3431}.

\bibitem[CF05]{MR2177713}
Gheorghe Craciun and Martin Feinberg, \emph{Multiple equilibria in complex
  chemical reaction networks. {I}. {T}he injectivity property}, SIAM J. Appl.
  Math. \textbf{65} (2005), no.~5, 1526--1546 (electronic). \MR{MR2177713
  (2006g:92075)}

\bibitem[Cha94]{bC94}
Brian Charlesworth, \emph{Evolution in age-structured populations}, Cambridge
  University Press, Cambridge, 1994.

\bibitem[Cha01]{bC01}
\bysame, \emph{Patterns of age-specific means and genetic variances of
  mortality rates predicted by the mutation-accumulation theory of ageing}, J.
  Theor. Biol. \textbf{210} (2001), no.~1, 47--65.

\bibitem[CLH99]{MR1754911}
VijaySekhar Chellaboina, Alexander Leonessa, and Wassim~M. Haddad,
  \emph{Generalized {L}yapunov and invariant set theorems for nonlinear
  dynamical systems}, Systems Control Lett. \textbf{38} (1999), no.~4-5,
  289--295. \MR{MR1754911 (2001h:34069)}

\bibitem[Daf78]{MR513814}
C.~M. Dafermos, \emph{Asymptotic behavior of solutions of evolution equations},
  Nonlinear evolution equations (Proc. Sympos., Univ. Wisconsin, Madison, Wis.,
  1977), Publ. Math. Res. Center Univ. Wisconsin, vol.~40, Academic Press, New
  York, 1978, pp.~103--123. \MR{MR513814 (80i:35019)}

\bibitem[Daw99]{Daw99}
Kevin~J. Dawson, \emph{The dynamics of infinitesimally rare alleles, applied to
  the evolution of mutation rates and the expression of deleterious mutations},
  Theor. Popul. Biol. \textbf{55} (1999), 1--22.

\bibitem[DS88]{MR1009162}
Nelson Dunford and Jacob~T. Schwartz, \emph{Linear operators. {P}art {I}},
  Wiley Classics Library, John Wiley \& Sons Inc., New York, 1988, General
  theory, With the assistance of William G. Bade and Robert G. Bartle, Reprint
  of the 1958 original, A Wiley-Interscience Publication. \MR{MR1009162
  (90g:47001a)}

\bibitem[EK86]{MR838085}
Stewart~N. Ethier and Thomas~G. Kurtz, \emph{Markov processes}, Wiley Series in
  Probability and Mathematical Statistics: Probability and Mathematical
  Statistics, John Wiley \& Sons Inc., New York, 1986, Characterization and
  convergence. \MR{MR838085 (88a:60130)}

\bibitem[Esh71]{MR0356905}
Ilan Eshel, \emph{On evolution in a population with an infinite number of
  types}, Theoret. Population Biology \textbf{2} (1971), 209--236.
  \MR{MR0356905 (50 \#9373)}

\bibitem[ESW06]{EvStWa06}
Steven~N. Evans, David Steinsaltz, and Kenneth~W. Wachter, \emph{A
  mutation-selection model for general genotypes with recombination}, 2006,
  Preprint, available at {\tt http://arxiv.org/abs/q-bio/0609046}.

\bibitem[ET77]{MR0682090}
Warren~J. Ewens and Glenys Thomson, \emph{Properties of equilibria in
  multi-locus genetic systems}, Genetics \textbf{87} (1977), no.~4, 807--819.
  \MR{MR0682090 (58 \#33112)}

\bibitem[Ewe04]{MR2026891}
Warren~J. Ewens, \emph{Mathematical population genetics. {I}}, second ed.,
  Interdisciplinary Applied Mathematics, vol.~27, Springer-Verlag, New York,
  2004, Theoretical introduction. \MR{MR2026891 (2004k:92001)}

\bibitem[FK00]{FiKi00}
Caleb~E. Finch and Thomas B.~L. Kirkwood, \emph{Chance, development, and
  aging}, Oxford University Press, 2000.

\bibitem[Ham66]{wH66}
W.~D. Hamilton, \emph{The moulding of senescence by natural selection}, J.
  Theor. Biol. \textbf{12} (1966), 12--45.

\bibitem[Hof85]{MR821683}
Josef Hofbauer, \emph{The selection mutation equation}, J. Math. Biol.
  \textbf{23} (1985), no.~1, 41--53. \MR{MR821683 (87d:92028)}

\bibitem[Hol95]{Hol95}
Robin Holliday, \emph{Understanding ageing}, Cambridge University Press, 1995.

\bibitem[Kin78]{MR0465272}
J.~F.~C. Kingman, \emph{A simple model for the balance between selection and
  mutation}, J. Appl. Probability \textbf{15} (1978), no.~1, 1--12.
  \MR{MR0465272 (57 \#5177)}

\bibitem[KJB02]{KiJoBa02}
Mark Kirkpatrick, Toby Johnson, and Nick Barton, \emph{General models of
  multilocus evolution}, Genetics \textbf{161} (2002), 1727--1750.

\bibitem[KM66]{KiMa66}
Motoo Kimura and Takeo Maruyama, \emph{The mutational load with epistatic gene
  interactions in fitness}, Genetics \textbf{54} (1966), 1337--1351.

\bibitem[Kon82]{Kon82}
A.S. Kondrashov, \emph{Selection against harmful mutations in large sexual and
  asexual populations}, Genet. Res. \textbf{40} (1982), 325--332.

\bibitem[LR04]{MR2104692}
Hartmut Logemann and Eugene~P. Ryan, \emph{Asymptotic behaviour of nonlinear
  systems}, Amer. Math. Monthly \textbf{111} (2004), no.~10, 864--889.
  \MR{MR2104692 (2005h:34132)}

\bibitem[Med52]{pM52}
Peter Medawar, \emph{An unsolved problem in biology: An inaugural lecture
  delivered at {U}niversity {C}ollege, {L}ondon, 6 {D}ecember, 1951}, H. K.
  Lewis and Co., London, 1952.

\bibitem[Nev75]{MR0402915}
J.~Neveu, \emph{Discrete-parameter martingales}, revised ed., North-Holland
  Publishing Co., Amsterdam, 1975, Translated from the French by T. P. Speed,
  North-Holland Mathematical Library, Vol. 10. \MR{MR0402915 (53 \#6729)}

\bibitem[SEW05]{StEvWa05}
David Steinsaltz, Steven~N. Evans, and Kenneth~W. Wachter, \emph{A generalized
  model of mutation-selection balance with applications to aging}, Adv. Appl.
  Math. \textbf{35} (2005), 16--33.

\bibitem[SH92]{SaHa92}
Stanley~A. Sawyer and Daniel~L. Hartl, \emph{Population genetics of
  polymorphism and divergence}, Genetics \textbf{132} (1992), 1161--1176.

\bibitem[TB90]{TuBa90}
Michael Turelli and N.H. Barton, \emph{Dynamics of polygenic characters under
  selection}, Theor. Popul. Biol. \textbf{38} (1990), 1--57.

\bibitem[WBG98]{MR1657856}
Holger Wagner, Ellen Baake, and Thomas Gerisch, \emph{Ising quantum chain and
  sequence evolution}, J. Statist. Phys. \textbf{92} (1998), no.~5-6,
  1017--1052. \MR{MR1657856 (2000e:92017)}

\bibitem[WES08]{0807.0483}
Kenneth~W. Wachter, Steven~N. Evans, and David~R. Steinsaltz, \emph{{The
  age-specific forces of natural selection and walls of death}}, 2008,
  Available at {\tt http://arxiv.org/abs/0807.0483}.

\bibitem[Wil57]{gW57}
George~C. Williams, \emph{Pleiotropy, natural selection, and the evolution of
  senescence}, Evolution \textbf{11} (1957), 398--411.

\bibitem[WSE08]{0808.3622}
Kenneth~W. Wachter, David~R. Steinsaltz, and Steven~N. Evans, \emph{{Vital
  rates from the action of mutation accumulation}}, 2008, Available at {\tt
  http://arxiv.org/abs/0808.3622}.

\end{thebibliography}
\end{document}